\documentclass[iop,apj
        ,twocolappendix
]{emulateapj}

  \usepackage{graphicx,color}% Include figure files
  \usepackage{natbib}
  \usepackage[normalem]{ulem}    %..  for strikeout
  \usepackage{amsmath,amssymb}
  \usepackage{float}
  \usepackage{setspace}
  \usepackage{gensymb}
   \newcommand{\vct}[1] {\ensuremath{\boldsymbol{#1}}}   %..    bold italic
    \newcommand{\vb} {\vct{b}}
    \newcommand{\vk} {\vct{k}}
    \newcommand{\vu} {\vct{u}}
    \newcommand{\vB} {\vct{B}}
    \newcommand{\vV} {\vct{V}}

  \newcommand{\tauNL} {\tau_\text{nl}}
  \newcommand{\di}    {d_\text{i}}
  \newcommand{\Omci}  {\Omega_\text{ci}}

  \newcommand{\avg}[1] {\ensuremath{ \langle {#1} \rangle }}
  \newcommand{\kmpssqd} {km$^2$/s$^{2}$}

\slugcomment{Draft Ver 2.2,  22-Apr-2014}

\shorttitle{Nonlinear and linear timescales near kinetic scales}

\shortauthors{Matthaeus et al.}
\begin{document}

\title{Nonlinear and linear timescales near kinetic scales
in solar wind turbulence}

\author{W.  H.  Matthaeus,$     ^{1\dagger}$,
         S.    Oughton,$    ^2$
         K.  T.  Osman,$      ^3$
         S.     Servidio,$  ^4$
         M.     Wan,$          ^1$
         S.  P.  Gary,$      ^5$
         M.  A.  Shay,$         ^1$
         F.     Valentini,$ ^4$
         V.  Roytershteyn,$  ^6$
         H.     Karimabadi,$   ^6$
        \&
         S.  C.  Chapman$     ^{3,7}$}
\affil{$^{1}$
Department of Physics and Astronomy,
University of Delaware, Delaware 19716, USA\\
$^{2}$
Department of Mathematics, University of Waikato, Hamilton, NZ\\
$^3$Centre for Fusion, Space and Astrophysics, University of Warwick, Coventry CV4 7AL, United Kingdom\\
$^4$
Dipartimento di Fisica, Universit\`a della Calabria, I-87036 Cosenza, Italy\\
$^5$
Space Sciences Institute, Boulder, CO 80301\\
$^6$ Sciberquest, Inc., Del Mar, California 92014, USA
\\
$^7$ Max Planck Institute for the Physics of Complex Systems, 01187 Dresden, Germany}
\email{$^{\dagger}$whm@udel.edu}

\begin{abstract}
The application of linear kinetic
treatments to plasma waves, damping, and instability
requires favorable inequalities
between the associated linear timescales
and timescales for nonlinear (e.g., turbulence) evolution.
In the solar wind these two types of timescales may be directly
compared using standard Kolmogorov-style analysis
and observational data.
% enables a direct comparison of
% nonlinear turbulence timescales
%and kinetic timescales in the solar wind.
%  including at scales of order the ion inertial scale, $ \di $.
The estimated local nonlinear magnetohydrodynamic cascade times,
evaluated as relevant kinetic scales are approached,
remain slower than the cyclotron period,
but comparable to, or faster than,
the typical timescales of
instabilities,
anisotropic waves, and wave damping.
The variation with length scale of the
turbulence timescales is supported by observations and simulations.
On  this basis
the use of linear theory---which assumes
constant parameters to calculate the
associated kinetic rates---may be questioned.
It is suggested that the product of proton gyrofrequency and
nonlinear time at the ion gyroscales provides a simple measure
of turbulence influence on proton kinetic behavior.
\end{abstract}

%----------------------------------------------------------------

        \section{Introduction}
        \label{sec:intro}

%% \sout{some rubbish}
%% \DEL{other rubbish}

Plasma physics often employs simplified
frameworks to explain properties of observed
plasmas in solar, space, and astrophysics.
Prominent among these
is the large class of calculations based
upon
linearization of a Vlasov description
about a uniform equilibrium magnetized state.
A rich variety of normal modes and wave damping rates
emerge, even when each plasma species possesses
simple properties such as Maxwellian distributions
with isotropic temperatures.
The dependence of these modes on species
plasma betas
and other dimensionless parameters
is a familiar and important topic in space plasma physics
and astrophysics.
More complex distribution functions
that admit temperature anisotropy or beams
are familiar in low collisionality solar wind,
accretion disks, and galaxy clusters
        \citep[e.g.,][]{SharmaEA07,SchekochihinEA10,
                        KunzEA11,RiquelmeEA12}.
These features provide free energy for
families of instabilities
such as the firehose, Alfv\'en ion cyclotron, and mirror
mode instabilities
        \citep[e.g.,][]{Gary}.
The characteristic
timescales (or reciprocal frequencies)
of the relevant
linear Vlasov modes
typically extend over a very wide range.

Intriguingly, many plasmas of interest---including
        the solar wind, corona, and interstellar medium---also
exhibit properties of a turbulence cascade
extending
from larger magnetohydrodynamic (MHD) to smaller kinetic spatial
scales.
Cascades also can extend over a wide range of timescales.
The present paper compares
        \emph{linear Vlasov timescales}
and     \emph{nonlinear turbulence timescales}
as the associated length scales approach the
transition between MHD and kinetic regimes.
We have in mind
the specific case of the solar
wind, for which it is possible to
inform the discussion using analytical
estimates, simulations,
and direct observational analysis.
We will conclude that
nonlinear and linear inverse
timescales can be comparable,
with frequencies of order of $1/100$ to $1/10$
the proton gyrofrequency, for the oblique wavevectors thought to dominate solar wind
fluctuations.
Therefore caution is required in applying the static
equilibrium assumptions underlying much of
linear theory.
Finally we close with a
suggestion for a simple dimensionless
measure of the degree of
turbulence cascade effects on kinetic processes.
Note that throughout the presentation we avoid
committing to a specific dynamical model of
the spectrum, such as Reduced MHD, two-dimensional MHD
or Goldreich--Sridhar theory
\citep{Mont82-strauss,ShebalinEA83,OughtonEA94,GoldreichSridhar95,ZhouEA04},
in order to maintain as
broad a context as possible;
however some issues related to
anisotropic spectral models are discussed
in Appendix~\ref{app:A}.

%------------------------------------------------------------------

        \section{Timescales in Linear Vlasov Plasma}
        \label{sec:linear}

%%Linear Vlasov theory
%gives rise to a variety of
%waves, damping rates, and instabilities
%of relevance
%in the corona and interplanetary medium.
From a technical perspective,
linearization leads to small amplitude
solutions having exponential behavior
        $\exp{(i \omega_c t)}$
with complex frequencies 
        $ \omega_c  =   \omega + i \gamma$
consisting of real frequency    $ \omega$
and a growth    ($ \gamma > 0 $)
or damping rate ($ \gamma < 0 $).
 In general,
        $ \omega $
and
        $ \gamma $
are functions of the wavevector   $ \vk $,
not just its magnitude            $ k =  | \vk | $.
Linearization
about a uniform state
yields normal modes of the plasma.
These are generally
transient
        \citep{Barnes79a},
but some have  small damping rates
        $ \gamma < 0$
with    $ |\gamma| / |\omega| \ll 1 $.
Besides damped waves,
relevant
instabilities
are studied in linear Vlasov calculations
by perturbing about
a simple plasma configuration
        \citep[e.g.,][]{Gary}.
A typical unstable equilibrium
might have uniform density and
magnetic field
        $\vB_0$,
with free energy supplied by an anisotropic
particle distribution function.

To discuss a range of relevant
timescales for normal modes,
it is convenient to adopt a
normalization that expresses timescales in units of
the proton gyrofrequency
        $ \Omci = eB/mc$,
in terms of a characteristic magnetic field strength $B$,
and length in units of
the ion inertial scale
        $ \di   =       c   / \omega_\text{pi}
                \equiv  V_A / \Omci $,
where   $ \omega_\text{pi}$
is the plasma frequency,
        $ c $
the light speed,
        $ \rho $
the mass density,
and
        $ V_A = B/\sqrt{4 \pi \rho} $
the Alfv\'en speed.
Kinetic scales will be
indicated when
        $ k \di \sim 1$
or greater,
while short timescales, $\tau( \vk)$,
are indicated by
        $ \Omci \tau( \vk) \sim 1$
or less.
The actual values of the frequencies associated
with waves and instabilities are obtained either by numerical
solutions of the full dispersion equation
        \citep{Gary,LysakLotko96}
or through analytic approximations
        \citep{Hollweg99-kaw}.

When the relatively low-frequency
MHD waves---Alfv\'en and fast and slow magnetosonic waves---are
extended to a kinetic description,
one finds
in linear Vlasov theory that
the magnetosonic waves
are much more
heavily damped
        \citep{Barnes66,Barnes68,Barnes69}
than the Alfv\'en mode for relevant parameters.
This damping is often invoked as
a basic physical explanation for frequently
observed fluctuations that resemble
the Alfv\'en waves in the inner heliosphere
        \citep{BelcherDavis71,RobertsEA87b}.
%As kinetic scales are approached
%the waves become dispersive and there is a general tendency for
%damping rates to increase.

Considerable effort has been devoted to describing
normal modes that may be present in the kinetic range
of solar wind turbulence, where
there is a well-known observed dominance of
quasi-two dimensional wavevectors
        \citep{MattEA90, LeamonEA00,OsmanHorbury07};
that is, perpendicular wavenumber
        $ k_\perp = |\vk \times \hat{\vB}_0|  \gg  $
parallel wavenumber
        $ k_\parallel = | \vk \cdot \hat{\vB}_0 |$.
In the following sections
we
therefore emphasize discussion of the
properties of oblique
fluctuations that are likely to make up
a substantial fraction of the solar wind
fluctuation spectrum.

\begin{figure}
\begin{centering}
\includegraphics[width=.99\columnwidth]{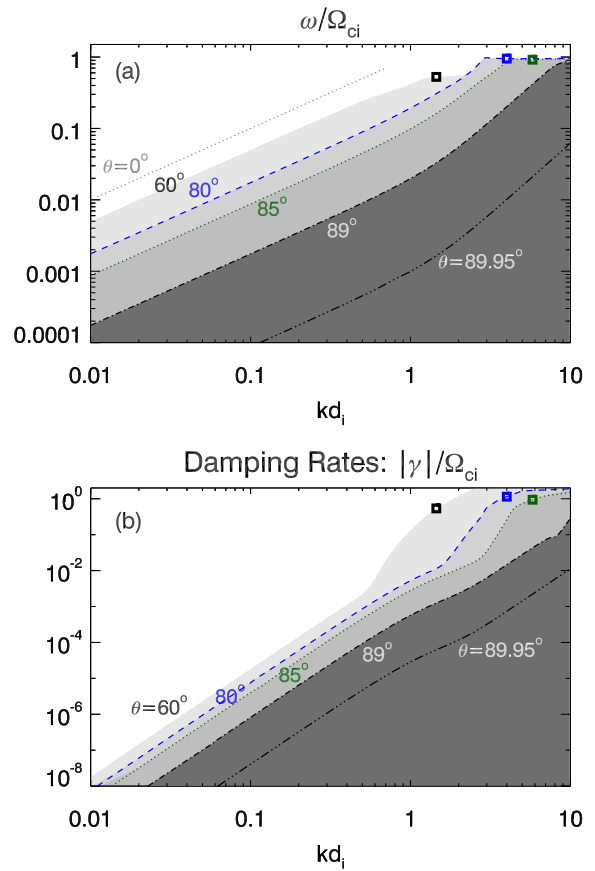}
  \caption{Illustration of the range of wave frequencies
obtained from linear Vlasov theory using
    typical solar wind parameters,
emphasizing oblique and quasi-two dimensional
wavevectors with angles to the
mean magnetic field in the range
        $ 60^\circ  <  \theta  < 90^\circ $, 
which are commonly thought to be most
relevant for solar wind turbulence cascade.
More oblique regions are shaded in a darker hue.
    (a) Real frequencies of Alfv\'en waves and KAWs.
    The curves at specific $ \theta $ are for a
        $ \beta_\text{p} = 1 $ plasma.
    (b) Wave damping rates, $|\gamma|$, for the same cases as
    in panel (a).
    Note that in the KAW regions there are cases where
                $ |\gamma( \vk)| > \omega( \vk) $
    and thus damping is too strong for waves to be properly excited.
    The first value of $ k \di $ where this occurs, at each $\theta$,
    is indicated by the square symbol.
}
 \label{fig:waves}
\end{centering}
\end{figure}

Within this class
a popular
choice is the oblique
kinetic Alfv\'en wave (KAW)
        \citep{Hollweg99-kaw,LeamonEA99}
with wave frequencies
        $ \omega < \Omci$
low compared to the
proton cyclotron frequency
        $ \Omci$.
Recent observations also suggest that
such low-frequency modes are energetically
most relevant in the solar wind between the ion and
electron inertial scales
        \citep{BaleEA05,HowesEA06,
        AlexandrovaEA09,SahraouiEA10-subproton}.
For this reason we will focus here on
wave properties approaching
and near        $k \di = 1$, 
and on wavevectors mainly in the oblique and quasi-two dimensional
range of angles to the mean field, that is
        $ 60^\circ < \theta < 90^\circ $.
Figures~\ref{fig:waves},
        \ref{fig:sims},
and
        \ref{fig:swdata}
portray this emphasis on oblique wavevectors by progressively
shading the linear results in the more oblique range of angles.

Higher frequency
waves may also be present, such as whistlers
        \citep{ChangEA13}
or      Alfv\'en  ion cyclotron (AIC) waves
with quasi-parallel wavevectors,
although these are generally thought
to occur at  a relatively lower amplitude
and higher frequency.
AIC modes are particularly relevant in
models involving pitch angle scattering
        \citep{IsenbergVasquez11}.

The observed frequencies and
wavevectors of fluctuations near ion kinetic scales
have been analyzed in terms of
linear wave theory
        \citep{SahraouiEA10-subproton,SahraouiEA12,RobertsEA13}.
Due to the ambiguities inherent in these analyses,
the main conclusion that can be drawn is that the
observed fluctuations in about a decade of scale near
        $ k \di \sim 1$
are ``low-frequency''
and are consistent
with a dominant contribution
of kinetic Alfv\'en waves with wavevectors lying in the
range of
        85\degree--89\degree\
of the mean magnetic field.
Interpreted as waves,
such fluctuations have frequencies roughly in the
range
        $ \omega / \Omci \sim 10^{-1}$
to      $ 10^{-2}$.
See
        Fig.~\ref{fig:waves}a.
At relatively short wavelengths ($ k \di  \gg  1 $)
and relatively high frequencies
        ($\omega/\Omci  \gg  1$ ),
whistler waves can propagate with
relatively weak damping at directions both parallel and oblique to the
background magnetic field
        \citep[e.g.,][]{GaryEA08}.
However,  there is substantial current debate as to whether such modes
make a significant contribution to the short-wavelength turbulent
spectra observed in the solar wind.
%---------------------------------------------------------
%---------------------------------------------------------

The \emph{damping} rates of normal modes such as KAWs are
also relevant, for example in some
theories of solar wind heating
that invoke a balance between linear damping
and cascade rate
        \citep{Barnes69,LeamonEA99,HowesEA08}.
Linear damping rates have been computed and tabulated for
both nearly parallel wavevectors
        \citep{GaryBorovsky04}
as well as highly oblique
orientations
        \citep{LeamonEA99, SahraouiEA10-subproton}.
Damping rates for modes with wavenumber
close to ion kinetic scales
        $ k \di \sim 1 $   (within a decade or so)
are frequently found to
be of order
        $ \gamma / \Omci   \sim 10^{-1}$
to      $ 10^{-2}$
        \citep{LysakLotko96,LeamonEA99,GaryBorovsky04}.
This
characteristic range of
damping rates
is found
for  a reasonably wide range of electron plasma beta
and for ratios of ion to electron temperatures
from zero to ten
        \citep{LysakLotko96}.
Depending upon parameters and angle of the wavevector to the mean
magnetic field, this damping rate may vary considerably.
   Figure~\ref{fig:waves}b
displays damping rates
for Alfv\'en waves and KAWs
in an electron--proton plasma
with proton plasma beta
        $ \beta_\text{p} = 1 $.
Note that for
        $ k \di \gtrsim 1 $,
putative KAWs at oblique angles have
        $ |\gamma(\vk)| / \omega( \vk) > 1 $
and thus are so strongly damped that it is difficult to excite them.

% although for KAWs one generally does not find damping rates
% of order one
%         ($ \gamma / \Omci \sim 1 $)
% until deep in the kinetic range at
% higher wavenumber
%         $ k \di \gg 1 $.

%---------------------------------------------------------
\begin{figure}
\begin{centering}
  \includegraphics[width=.99\columnwidth]{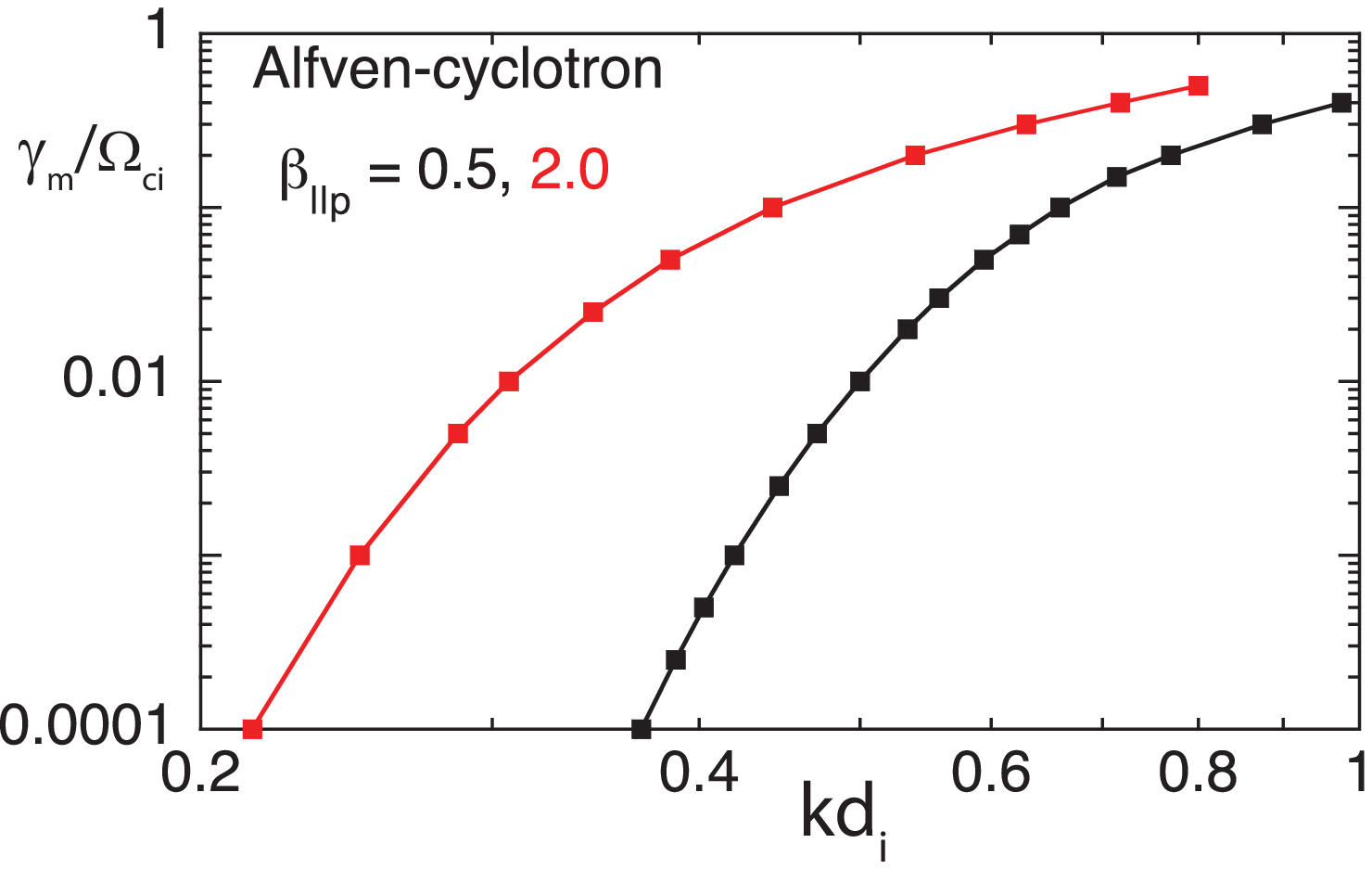}
  \includegraphics[width=.99\columnwidth]{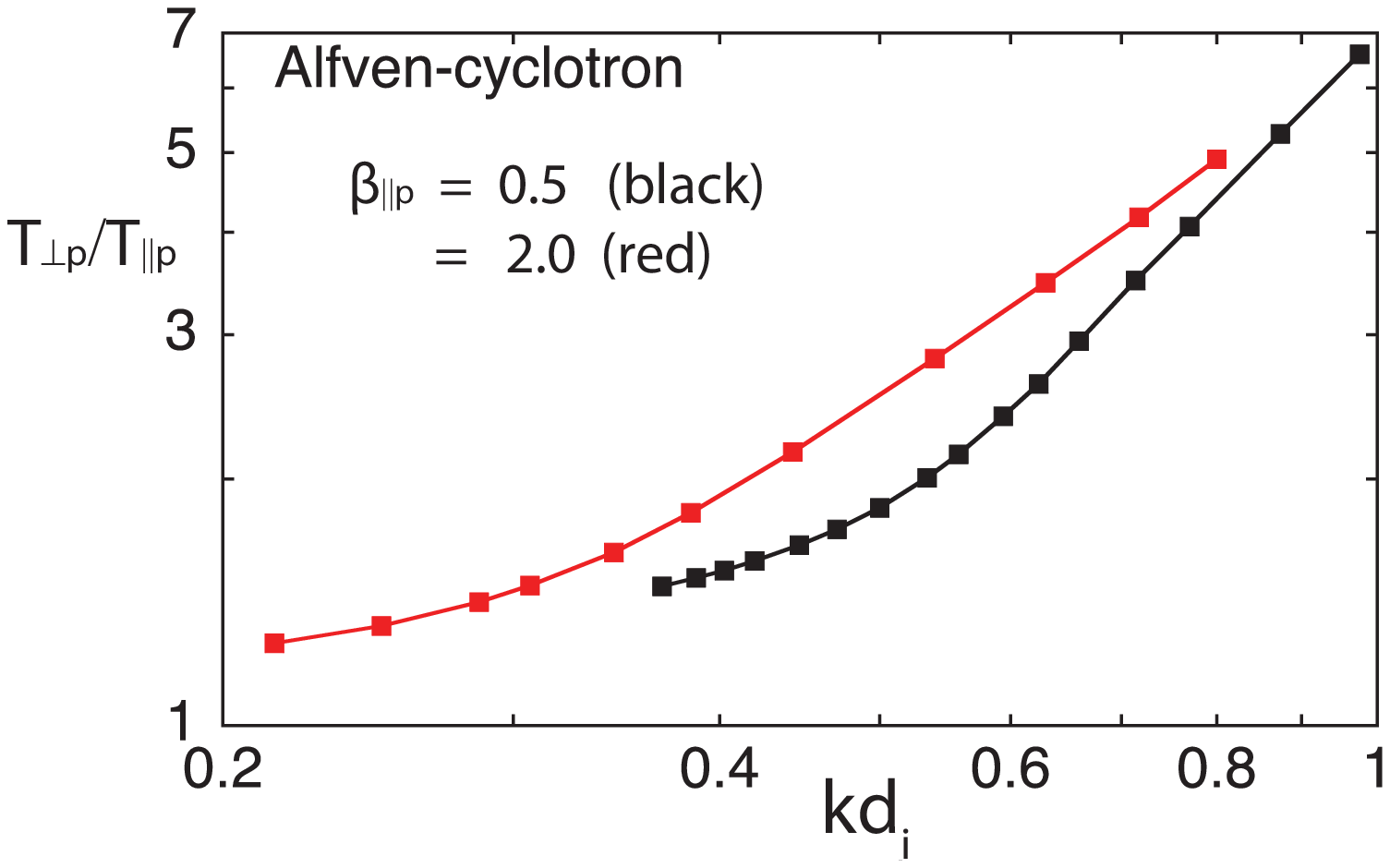}
  \caption{(a)
   Maximum growth rates $\gamma_m$
   normalized to ion cyclotron frequency  $ \Omci$
   vs.\  wavenumber normalized to ion inertial length $\di$,
   for the Alfv\'en cyclotron
   instability, for two values of $\beta_{\parallel\text{p}}$ typical
   of the solar wind.
    (b) Corresponding proton temperature anisotropies for the cases
    shown in panel (a).
Note that plasma with
values of $\beta_\parallel$ and $T_{\perp p}/T_{\parallel p}$
such that the growth rate 
        $ > 0.05 \Omci $
are rarely observed in the solar wind.
}
 \label{fig:instabilities}
\end{centering}
\end{figure}
%---------------------------------------------------------

%new section from Peter

If the ion and electron velocity distributions are nearly thermal,
that is, approximately Maxwellian, kinetic linear dispersion theory
predicts that the normal modes of the plasma are undamped or weakly
damped.
However, if a species velocity distribution is sufficiently
anisotropic, dispersion theory as well as kinetic simulations show
that normal modes grow in time, leading to instability.   Various
anisotropies drive a variety of instabilities [Gary, 1993]; a source
of free energy often observed in the solar wind is the
        $ T_{\perp \text{p}} / T_{\parallel \text{p}} > 1 $
proton temperature anisotropy.

Under typical
solar wind conditions, there are two distinct growing modes which
arise as a result of this anisotropy:
the proton mirror instability and the
Alfv\'en-cyclotron instability.
The former has zero real frequency in a homogeneous plasma, has
maximum growth rate
        $ \gamma_\text{m} $
at relatively oblique propagation
  ($ 0 <  k_{\parallel}  <  k_{\perp} $),
and is favored at relatively high values of
        $ \beta_{\parallel \text{p}} $.
The latter mode satisfies
        $ 0 < \omega < \Omci $,
has maximum growth rate at
        $\vk \times \vB_0 = \vct{0} $,
and is preferentially excited if
        $ \beta_{\parallel \text{p}}  < 1 $.
As the proton temperature
anisotropy is increased, Figure 2a-b, illustrating the
typical solar wind parameter range of
        $ 0.5 < \beta_{\parallel \text{p}} < 2.0 $,
show that linear dispersion theory
predicts that both
        $\gamma_\text{m}$
and the corresponding wavenumber
        $k \di$
also increase in magnitude.
Spacecraft observations
   \citep{HellingerEA06,MatteiniEA07,BaleEA09,MarucaEA11}
show that scattering by enhanced
fluctuations from instabilities acts to constrain proton anisotropies.
The typical extremal anisotropy values correspond
to relatively weak growth rates, that is,
        $ 10^{-3}   < \gamma_\text{m} /\Omci \lesssim 0.05 $.
   Figure~\ref{fig:instabilities}a
shows that this growth rate corresponds to
        $ 0.25  < k \di < 0.6 $.
This implies that the fluctuations that may be produced by
 proton-driven instabilities would be expected
to have
their maximum amplitudes at wavelengths
near (or slightly larger than)
the ion inertial scale, which typically
marks the end of the inertial range
spectrum, as discussed below.

Summarizing these linear theory results,
near the scales associated
with onset of the kinetic physics range
        ($ k \di \sim 1 $)
there are highly relevant waves with frequencies
on the order of a few tenths of the cyclotron
frequency or less,
and associated damping rates of similar magnitude
that may contribute to dissipation.
Likewise, temperature anisotropy-driven instabilities
at limiting parameters in the solar wind
are reported to have typical growth rates
that are also less than a tenth of the cyclotron frequency.
%------------------------------------------------------------------

        \section{Nonlinear Cascade Timescales}
        \label{sec:nonlinear}

Plasma dynamics at finite amplitude
permit nonlinear couplings that are
contemporaneous with linear processes.
Sufficiently strong nonlinearity
drives
a cascade that
potentially influences dynamics
across decades of scale.
In the solar wind this picture is supported
by observation of powerlaw
energy spectra, evolving Alfv\'en ratio and
cross helicity, and the distribution of plasma
heating
        \citep[e.g.,][]{DobrowolnyEA80-prl,%
                        GoldsteinEA95,TuMarsch95,MattVelli11}.
The cascade is also evidenced directly,
by observation of third-order statistics
        \citep{Sorriso-ValvoEA07,MarinoEA08,MacBrideEA08}.

Accompanying broadband spatial structure,
a wide range of timescales
also characterizes the
cascade.\footnote{It is important
        to distinguish Eulerian timescales
        and Lagrangian timescales.     The former
        may include fast timescales associated
        with sweeping and unidirectional
        wave propagation, which do not induce spectral
        cascade
          \citep{ChenKraichnan89,ServidioEA11-epl}.}
These
nonlinear timescales
generally become smaller at smaller scales.
This speed-up is important as it
is responsible for the
tendency of turbulence
to attain quasi-universal small-scale statistical equilibria
        \citep{BatchelorTHT}.
The question at hand is whether,
as the kinetic plasma range is approached in scale
(and at appropriate oblique angles to the mean magnetic field),
the nonlinear timescales are competitive with
timescales emerging from the linear processes
summarized in the previous section.
This comparison relates to the balance
between cascade activity---mainly mediated
    by the nonlinear time
   (see below, and Appendix~\ref{app:A})---and
linear kinetic effects
that occur independently at each
wavevector without regard for cross-scale
couplings.

For context, let us review
the standard Kolmogorov phenomenology for steady-state isotropic
hydrodynamics
        \citep[e.g.,][]{Frisch},
focusing on the scale-dependence of the nonlinear timescales.
Denote the (rms) turbulence amplitude as $u$
and the outer (or energy-containing) scale as $L$.
The global cascade rate $\epsilon$ and the scale-dependent version
        $\epsilon_\ell$
are constrained as
\begin{equation}
  \epsilon
        \sim
  \frac{u^3}{L}
        \sim
  \epsilon_\ell
        \sim
  \frac{u_\ell^3}{\ell}
        \sim
  \frac{u_\ell^2}{\tau_\ell} ,
 \label{epsilons}
\end{equation}
in terms of inertial range scale        $ \ell$
and
longitudinal velocity increment         $ u_\ell$.
The last relation serves to define the
scale-dependent nonlinear time
\begin{equation}
  \tau_\ell
        =
  \frac{\ell}{u_\ell}
        \sim
  \frac{\ell^{2/3}} {\epsilon^{1/3}}.
\label{tau_ell}
\end{equation}
The equivalent development in Fourier wavenumber   $ k$
characterizes a steady cascade,
local in wavenumber, as
\begin{equation}
   \epsilon
        \sim
   \epsilon_k
        \sim
   \frac{u_k^2}{\tauNL(k)}.
\label{epsilonk}
\end{equation}
Here the amplitude
of fluctuations at scale $1/k$ is
        $ u_k = \sqrt{ k {\cal E}(k) } $
 [dropping $ O( 1) $ factors]
and the nonlinear time at wavenumber $k$ is
\begin{equation}
  \tauNL(k)
        =
   \frac{1} {{ k u_k }}
        \sim
   \frac{L}{u}  \frac{1} {(kL)^{2/3}}.
\label{tauNLk}
\end{equation}
In writing
        Eq.~(\ref{tauNLk})
use is made of the
steady Kolmogorov omnidirectional spectrum:
        $ {\cal E}(k) \sim \epsilon^{2/3} k^{-5/3} $.

Extension of this result to
MHD is straightforward
        \citep[e.g.,][]{ZhouEA04}.
For nearly incompressible MHD
the relevant cascaded quantity
is the total energy per unit mass, essentially
        $ Z^2 = u^2 + b^2 $.
%%  \\ \comment{Dropped 1/2 off this
%%        (so $Z^2/2 =$ energy and $Z \equiv u $ when $b=0$)}
Here,
         $u$ and $b$ are the rms
fluctuations in velocity and magnetic field,
the latter measured in Alfv\'en speed units.
For the present illustration we
consider the simplest case in which,
for the inertial range of scales,  the
cross helicity is near zero
        (i.e.,  uncorrelated $\vu$ and $\vb$),
and $u^2$ and $b^2$ are
of the same order.\footnote{%
        For the finite cross helicity case, see, e.g.,
          \cite{DobrowolnyEA80-prl,ZhouEA04,MattEA04-Hc}.}
Then the
above arguments are readily reformulated in
terms of the total energy and
        $ Z_k = \sqrt{ k {\cal E} (k) } $,
the amplitude
near wavenumber $k$.\footnote{%
        $ {\cal E}(k) $ is now the omnidirectional spectrum for the
        total (kinetic plus magnetic) energy.}
The relevant nonlinear timescale for a Kolmogorov analysis
of MHD becomes
\begin{equation}
  \tauNL(k)
        =
   \frac{1} {k Z_k}
        \sim
   \frac{L}{Z}  \frac{1} {(kL)^{2/3}} ,
\label{MHDtime}
\end{equation}
in direct analogy to
        Eq.~(\ref{tauNLk}).

It is also straightforward
to introduce modifications in the
above reasoning to treat the
anisotropic perpendicular MHD
cascade that is obtained in the presence of
a strong imposed mean magnetic field
        \citep{ShebalinEA83,ZhouEA04}.
For an assumed
perpendicular
cascade, the familiar procedure
        \citep[e.g.,][]{GoldreichSridhar95}
is to simply interpret the scale        $ \ell$
and  the wavenumber                     $ k$
as the projection onto the
perpendicular plane.
Another potentially important timescale in describing
the cascade is the Alfv\'en
crossing time.
However for reasons outlined in
        Appendix~\ref{app:A},
we will base the following discussion 
 of cascade timescales
only on the nonlinear timescale.

Moving into the realm of kinetic plasma
dynamics, there is no generally
accepted formulation of a nonlinear timescale,
in contrast to the fluid regime.
  %%        \citep[e.g.,][]{ZhouEA04}.
However
we expect the same kind of
hydrodynamic advective and line-stretching nonlinearities
to be present in kinetic plasma.
For  $k \di \geq 1 $ 
there should also be effects of the Hall current and other
contributions to the generalized Ohm's Law, which would change
the estimate of the nonlinear timescale from
        $ \tauNL( k) \sim   1/ ( k Z_k) $ 
to something with a stronger dependence on $k$, such as 
        $ \tauNL( k) \to 1/( \di k^2 Z_k)$.
The introduction of new timescales
leads to the possibility of several different spectral scalings,
an effect familiar in Hall MHD or electron MHD
turbulence studies 
        (e.g., \citet{BiskampEA99,GaltierBuchlin07,AlexandrovaEA08-swind}).
Due to these complications 
(as well as the practical matter of the lack of high-frequency
velocity data), 
in the analysis below we will restrict
estimates to the simplest local MHD nonlinear timescale.
We expect that in the deep
kinetic regime the MHD
timescales will be upper bounds
for the actual nonlinear timescales.
However we will defer to a future study
a more careful and detailed treatment of the
        $ k \di > 1 $ 
nonlinear timescales.

With these caveats in mind,
in all cases below
we will estimate relevant nonlinear timescales
by adopting the formulation given
in
        Eq.~(\ref{MHDtime}).
%-----------------------------------------------------------
\begin{figure}
\begin{centering}
\includegraphics[width=0.99\columnwidth]{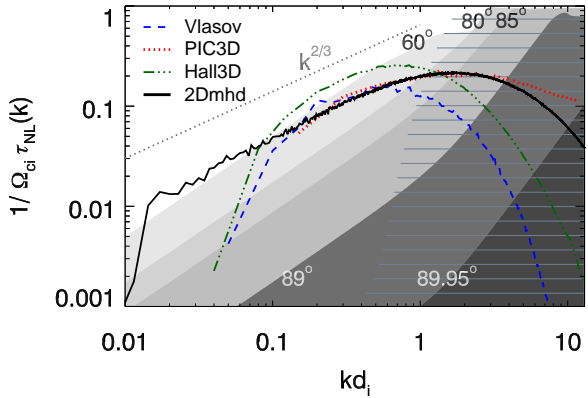}
\caption{Wavenumber dependence of the estimated
        nonlinear rates computed from Eq.~(\ref{MHDtime})
        for several types of simulations
        (see text).
        As discussed in the text, no attempt is made to strictly
        control the simulation parameters.
        Nevertheless there is substantial physical similarity---the
        nonlinear rates increase as the scales decrease towards 
        the dissipative or kinetic regime, and become comparable
        in magnitude to a tenth or more of the cyclotron
        frequency.
%         (2D MHD,
%         3D Hall MHD,
%         2.5D Eulerian Vlasov,
%         and 3D kinetic PIC).
         Also shown is a reference $ k^{2/3} $ line,
        associated with the scaling of
        $ 1/\tauNL( k) $
       for a steady Kolmogorov cascade.
        Shading corresponds to ranges
      of linear wave frequencies $\omega$,
      at increasingly oblique angles (darker shading).
%        of
%            60$^\text{o}$,
%            80$^\text{o}$,
%            85$^\text{o}$,
%            89$^\text{o}$,
%            and 89.95$^\text{o} $.
       Horizontal hatching indicates corresponding linear theory
       damping rates $\gamma$ (cf.\ Figure~\ref{fig:waves}).
}
 \label{fig:sims}
\end{centering}
\end{figure}
%-----------------------------------------------------------
The expected speed-up
of the nonlinear rate is seen explicitly in the variation of
        $ 1/\tauNL(k) \sim k^{2/3}$,
as illustrated in
        Fig.~\ref{fig:sims}.
Here we depict
an idealized inertial range behavior of  $ 1/\tauNL( k) $,
extending from the outer scale to the Kolmogorov
dissipation scale.
Also shown are scale-dependent nonlinear rates computed from several
types of simulation data,
using the inertial range formula Eq.~(\ref{MHDtime}).
Included are data from simulations of
        2D MHD,
        3D Hall MHD,
        3D PIC kinetic plasma
 \citep{RoytershteynEA14},
and     2.5D hybrid Eulerian Vlasov  \citep{ServidioEA12}.
In this selection of
simulations the initial conditions and parameters
have some similarities---rough equipartition of velocity
and magnetic field fluctuations,
minimal compressibility effects, equal viscosity and
resistivities when possible, etc.  However
the systems are not strictly controlled to be identical.   For example
the mean field strengths vary
 (including no mean field for the 2D MHD case), and
 parameters such as Hall parameter and mass ratio
may differ.
Data are taken at or near the time of peak mean-square current density.
Further simulation details are given in
        Appendix~\ref{app:B}.

Note that in calculating $\tauNL(k)$
we employ twice the magnetic (omnidirectional) energy spectrum, i.e.,
        $ Z_k = \sqrt{2 k  {\cal E}^b( k)} $,
with the factor of two accounting for the approximately equal kinetic
and magnetic contributions.
Although this approach is not necessary with simulation data, it
does facilitate later comparison with solar wind observations
        (Section~\ref{sec:comparison}),
for which the cadence of the plasma (velocity) data is often much
lower than that of the magnetic field measurements.

In presenting the simulation results,
it is convenient to normalize wavenumbers to
the ion inertial length,
        $ \di $.
For the kinetic cases $ \di $ is intrinsic in the numerical
formulation.
For (one-fluid) MHD simulations this is not so, and there we have
associated      $ \di $
with
        $  \sqrt{43} \eta $,
where $\eta$ is
the Kolmogorov dissipation scale computed from the simulation
and
        $ 43^2 \approx 1836 $
is the proton/electron mass ratio.
This approach places the dissipation scale at the geometric mean of
the ion and electron inertial scales.
Consequently, all results in Fig.~\ref{fig:sims}
are presented using the normalization
      $ [k \di, \, 1/(\Omci \tauNL(k)) ] $.

It is readily apparent that, in all cases, the nonlinear
rates increase with  $ k $, until a steepened dissipation range is encountered,
whether a well-defined inertial range is seen, or not.
In the various
simulations one always finds
that the slope of the nonlinear rate
passes through a region in which it is similar
to the Kolmogorov value.
As the dissipation range is approached,
the spectral density of energy decreases more rapidly and
the nonlinear rate levels off and usually
decreases at very small scales.
However in the upper inertial range---near
        $ k \di \lesssim 1 $---the
fastest
nonlinear rates are entering the regime
of kinetic rates since
        $ 1/ \Omci \tauNL( k) \approx 0.1$--0.2,
  i.e, \emph{not} $ \ll 1 $.
The nonlinear ``fluid''
timescales remain longer than the
proton gyroperiod
        $ \tau_\text{ci} = 2\pi/\Omci $.
However,
the figure also shows that the nonlinear timescales
in the crucial transition range between fluid and
kinetic scales remain  faster than essentially all
frequencies of highly oblique
        $ \theta > 80^\circ $ linear waves, 
and faster than the associated damping rates of these waves.
The nonlinear timescales are more than an order of magnitude
faster than the wave frequencies of extremely
oblique 
   ($ > 89^\circ $) 
fluctuations at   $ k \di = 1 $.
%------------------------------------------------------------------

        \section{Linear, Nonlinear, and Solar Wind
          Timescales}
        \label{sec:comparison}

Linear and nonlinear processes are
concurrent in a dynamic plasma
and comparison of their
characteristic timescales
is a useful basis for
discussing their relative effects.
For example, when the nonlinear timescales are
extremely long compared to the timescales
computed for linear processes, then one expects
those linear processes to occur without
immediate modification.
On the other hand, when nonlinear effects
occur over a timescale comparable to,
or shorter than,
those of linear effects,
one must pause to reconsider how these processes
interact with one another.

%-----------------------------------------------------------
\begin{figure}
\begin{centering}
\includegraphics[width=0.99\columnwidth]{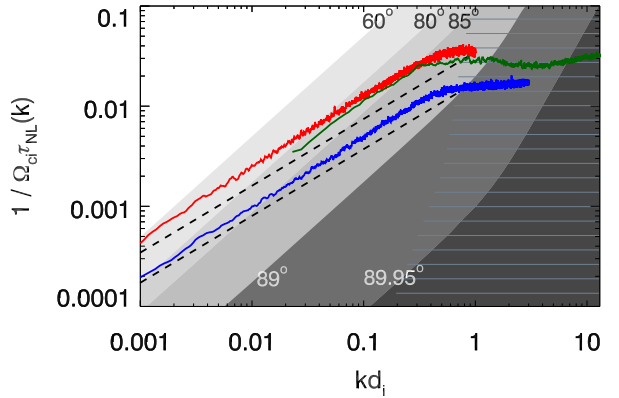}
\caption{Nonlinear rates
        (reciprocal nonlinear times) for
        two fast solar wind intervals (red, green)
        and a slow solar wind interval (blue),
        compared to theoretical nonlinear rates for some typical solar
        wind parameters (dashed);
         see text and Table~\ref{tab:sw-intervals}.
        Also indicated are linear Vlasov theory determinations for
        characteristic wave frequencies (shaded regions),
        and damping rates (horizontal hatching).
        For kinetic scales ($ k \di \sim 1 $ or greater), the
        observational nonlinear rates are comparable to the linear
        theory frequencies and rates associated with oblique angles.
        Shading delineates more oblique angle linear results,
    as in earlier figures.
}
 \label{fig:swdata}
\end{centering}
\end{figure}
%-----------------------------------------------------------
To affect such a comparison,
        Fig.~\ref{fig:swdata}
presents
a digest of frequencies from linear theory (shaded region),
as in Fig.~\ref{fig:sims},
along with
observationally determined nonlinear rates from three
%%        sample
solar wind intervals.
These observational estimates are based on
        $ E^b( k) $,
the reduced energy spectrum for the magnetic field,
since the reduced spectrum for the velocity is often
not available near the scales of interest here.
To correct for this, we assume approximate equipartition of kinetic
and magnetic energy at inertial range scales,
i.e., we employ
        $ 1 / \tauNL(k) = k \sqrt{2 k E^b(k)} $.
(Except that the omnidirectional spectrum has been replaced by the
reduced one, this is the same definition used for Figure~\ref{fig:sims}.)

% Also shown are model nonlinear rates
% for some typical solar wind parameters,
% calculated using
%% the  reciprocal of
%        Eq.~(\ref{MHDtime}).

%..  beta_p = n_p k_B T_p / [B^2 / (2mu_0)]
%------------------------------------------------------------------------------%
  \begin{table} [tbp]           %% top,bottom,page,here
    \caption{Some parameters of the solar wind intervals used in
             Figure~\ref{fig:swdata}.}
        \label{tab:sw-intervals}
    \begin{center}
    \begin{tabular} { l l| c c c } %% \hline
        Quantity &   & Wind & STEREO & Cluster \\
      \hline
        \avg{|V_{sw} |}        &  km/s     & 384  & 661  & 666 \\
        \avg{|V_A |}           &  km/s     & 70   & 65   & 60 \\
        \avg{|B |}             &  nT       & 7.4  & 3.7  & 4.5 \\
        \avg{|\delta V_A |^2}  &  \kmpssqd & 294  & 164  & 20 \\
        \avg{|\delta V_{sw} |^2}& \kmpssqd & 361  & 974  & 326 \\
%        \avg{ T_p }             & $10^5$\,K& 0.84 & 1.68 & 5.7 \\
        \avg{ n   }             & cm$^{-3}$& 5.3  & 1.5  & 2.6 \\
        $\di$                   & km       & 99   & 185  & 141 \\
        $\Omci$           & rad/s    & 0.71 & 0.36 & 0.43 \\
        $\beta_\text{p} $       &          & 0.3  & 0.6  & 2.5  \\
%         Time                    & UT       & Apr\,2000 & May\,2007 & Jan\,2007 \\
%                                 &            & 03\,09:00--
%                                              & 25\,00:00--
%                                              & 30\,00:00-- \\
%                                 &           & 06\,15:00
%                                             & 28\,02:39
%                                             & 30\,01:20 \\
        $B$ resolution          & Hz       & 11   & 8    & 450 \\
      \hline
    \end{tabular}
    \end{center}
  \end{table}
%%              left,center,right.    page 64 of LaTex book;
%       \multicolumn {# cols to cover} { l, c, or r} {text-string}
%       \cline {start_col-end_col}
%       \rule [-1.7ex]{0pt}{5.0ex}
%      \begin {tabular} { || p{5em} | p{8em} ||  }
%------------------------------------------------------------------------------%

%% paragraph from Kareem
We use 8\,Hz magnetic field measurements from the IMPACT instrument
        \citep{AcunaEA08,LuhmannEA08}
and 1\,min resolution proton plasma data
from the PLASTIC instrument
        \citep{GalvinEA08} onboard the STEREO
spacecraft in the ecliptic.   A total of nine STEREO intervals are
used, where all are in high-speed streams and contain no sector
crossings.   These intervals are identical to those used by
        \cite{Podesta09-SWaniso}.
Figure~\ref{fig:swdata} shows nonlinear rate
estimates (in red) from
        2007, May 25 00:00 to 28 02:39 UT, which
is typical of the stationary fast solar wind intervals used in this
study.
These are compared to data from slow solar wind intervals,
where we use 11\,Hz magnetic field measurements from the MFI
instrument
        \citep{LeppingEA95} and 3\,s resolution plasma data from
the 3DP instrument
        \citep{LinEA95} onboard the Wind spacecraft at  1\,AU.
%% Figure~\ref{fig:swdata} shows
Shown is a typical nonlinear rate
estimate (in blue), from a slow solar wind stream in
        2000, April 03 09:00 to 06 15:00 UT.
The data in these intervals has been truncated
from
         $ k d_{i} > 1$ for STEREO
and from $ k d_{i} > 3$ for Wind,
since
noise becomes important at these scales and leads to an artificial
flattening of the power spectral density.
For completeness and in
order to compare with kinetic simulations, we use high-frequency
measurements of the magnetic field fluctuations from the search-coil
(STAFF-SC)
        \citep{CornilleauWehrlinEA97}
and flux-gate magnetometers (FGM)
        \citep{BaloghEA97}
onboard the Cluster spacecraft quartet to probe
kinetic scales.
We have chosen an interval where both STAFF-SC and FGM are operating
in burst mode so that the smallest scales are
accessible.
Plasma data is obtained from the CIS HIA
        \citep{RemeEA97}
instrument on Cluster 1.
This is for a fast solar wind stream in
        2007, January 30 00:10 to 01:20 UT.
The estimates of the nonlinear rates are shown (in green) in
        Figure~\ref{fig:swdata}.
The analysis is restricted to frequencies lower
than 40\,Hz to maintain a signal to noise ratio no less than 10\,dB
        \citep{KiyaniEA09}.

Also shown in
        Fig.~\ref{fig:swdata}
are two theoretical curves,
computed using $ \Omci $ times  Eq.~(\ref{MHDtime}),
and employing average
solar wind-like parameters:
        $ V_A = 60  $\,km/s,
correlation scale of
        $ 10^6      $\,km,
and
        $ \di = 100 $\,km
        (i.e., density $\approx 5/$cm$ ^3 $).
The curves differ due to the choice of
squared fluctuation amplitude:
        $ Z^2 = 500 $\,km$^2$/s$^2$
and     $ Z^2 = 2000$\,km$^2$/s$^2$.
For clarity we do not repeat the several
simulation results in
        Fig.~\ref{fig:sims}.

The parameter space regions shaded in
        Fig.~\ref{fig:swdata}
correspond to same prominent
\emph{linear} processes: wave oscillation frequencies,
wave damping rates, and instability rates, as were
depicted in Fig.~\ref{fig:sims}.
These areas correspond to the
discussion in
        Section~\ref{sec:linear}
and the examples given in
        Fig.~\ref{fig:waves}.
Evidently, the nonlinear rates are comparable to the
linear ones near the onset of kinetic scales,
        $ k \di \sim 1 $
for sufficiently oblique spectral distributions of
energy, and especially for 
        $ \theta > 89^\circ $.
Note that the Wind interval in particular (slow wind case)
has a lower beta $0.3$ for which the higher frequency
dispersion relations should be shifted downwards
relative to the $\beta = 1$ shown here.

%-----=--------------------------------------------------------

        \section{Discussion:
Nonlinear effects in the kinetic regime}
        \label{sec:nl-kinetic}

It is apparent from
        Fig.~\ref{fig:swdata}
that
at length scales approaching $k \di = 1$ from above,
the rate of local-in-scale
nonlinear processes overtakes and then exceeds
a number of the kinetic plasma processes that
have received significant attention
regarding solar wind
plasma dispersion and dissipation in the kinetic range.
This is particularly true for highly oblique linear modes,
which, for emphasis,
are depicted using darker shading
in Figs.~\ref{fig:waves},
        \ref{fig:sims},
and     \ref{fig:swdata}.
This effect
may not have received sufficient consideration previously.
Its implications, however, may differ
subtly depending on which type of kinetic process is
under consideration.

For \emph{linear waves}, the influence of a fast
nonlinear cascade is expected to modify the
dynamical response of the system,
        i.e.,  the activity at a specified wavevector.
Instead of being a simple oscillator with a
characteristic frequency,  one now has
driving from larger scales, and damping by transfer
of energy to smaller scales.     So the problem becomes one
of a stochastic nonlinear oscillator, which may exhibit
behavior much different from simple harmonic motion.
For example, random scrambling of a wave phase,
even without energy change, causes a potentially
dramatic frequency broadening   \citep{vanKampen}.
Driving may also randomly change the
energy content of the wave.
Clearly, linear couplings will
remain present and under some conditions
linear wave properties may play an important role.

Turning to the
        \emph{damping rates},
the situation becomes somewhat different.
When the cascade rate at a particular scale
becomes larger than
the linear Vlasov damping rate $\gamma$,
then the latter may become increasingly irrelevant.
This may happen for example at a wavevector
        $ \vk $ 
when the nonlinear time
        $\tauNL (\vk)  \ll  \gamma( \vk)$.
In this case for quasi-steady cascade conditions,
most of the damping is due to
dissipation not at      $ \vk $,
but in other (smaller scale) fluctuations.
For example if the nonlinear cascade effect
is much faster than
linear damping of a particular KAW,
        %% kinetic Alfv\'en wave,
then the energy may be transferred to much
smaller scales and damped by other processes
(including possibly damping of whistlers).

Finally,
        \emph{instability calculations}
may need be modified to account for nonlinear rates that exceed
standard instability growth rates.
Indeed it would seem
that the problem of instability in a steady
        \emph{cascade}
becomes a perturbation about a driven dissipative
steady-state,
in contrast to one about an equilibrium.
It is unclear whether the growth rates and
other properties of usual
mirror mode,
AIC,
and firehose instabilities
        \citep{Gary,HellingerEA06}
will be changed substantially,
and further detailed work on specific cases will be required
to address this question.
It is interesting to note that in the parameter space
regions in which these instabilities are expected
to act, there is also accumulating evidence
of effects that might be attributable to
enhanced turbulence
        \citep{BaleEA09,
                OsmanEA12-kinetic,OsmanEA13,
                ServidioEA14-vlasov}.

The above considerations
suggest a natural measure of the degree to
which the local ion kinetic physics
is influenced by the nonlinear
MHD-scale cascade.
The relevant parameter appears to be
 \begin{equation}
 \Phi( \di) \equiv \Omci \tauNL (k \di = 1).
\label{eq:phi-di}
\end{equation}
We have in mind a solar wind plasma
with
        $ \beta_\text{p} \approx v_\text{th}^2 / V_A^2 \sim 1 $
for thermal speed
        $ v_\text{th} $.
For
systems with more widely ranging
        $ \beta_\text{p} $,
a more accurate indicator
may be
\begin{equation}
\Phi(\rho_i) = \Omci \tauNL( k \rho_i=1),
\label{eq:phi-rhoi}
\end{equation}
where   $ \rho_i$
is the thermal proton gyroradius.
Using the appropriate definition
(which may vary according to specific cases),
        $\Phi < 1 $
indicates that the kinetic physics
is strongly influenced by the MHD-scale cascade.
However, since many relevant
linear wave frequencies,
damping rates,
and instability growth rates in the solar wind lie in the
range
%%        $ \le 10^{-1} \Omci $,
        $ \lesssim \Omci /10 $,
there may be significant nonlinear influences
even when
        $ \Phi (\di)  < 10$ to  $100$.

For the idealized inertial range, the above
estimate of the
normalized nonlinear timescale at
        $ \di$
may be obtained  using
        Eq.~(\ref{MHDtime}):
\begin{equation}
       \Phi( \di)  \equiv
  \Omci \tauNL( \di)
        =
  \frac{L \Omci}{Z} \frac{\di^{2/3}}{L^{2/3}}
        =
  \frac{V_A}{Z} \left( \frac{L}{\di} \right )^{1/3},
\label{TNLatdi}
\end{equation}
a form that may be useful when
$\beta_\text{p} \sim 1$
and cross helicity $H_c \approx 0$.
This can readily be
generalized for other cases.
The latter
characterization should not be
applied at scales
smaller than    $ k \di \sim 1 $,
given that the
form of         $\tauNL( k)$ 
is likely different as discussed above.
However a criterion for significance of
nonlinear effects based on
the more general form
given in Eqs.~(\ref{eq:phi-di}) and
        (\ref{eq:phi-rhoi})  
may remain valid
even at         $ k \di > 1 $.

We may note the relationship between
        $ \Phi $
and a familiar measure of turbulence strength,
the effective Reynolds number,
which
may be estimated as
        $ R_\text{eff} =  ({L} / {\di})^{4/3} $.
For nominal
solar wind parameters,
with
        $ \beta_\text{p} = 1  $,
        $   L = 10^6 $\,km,
        $ \di = 100  $\,km,
and     $ Z/V_A$ in the range $1/2$ to $1$,
we find that
        $ \Phi( \di) = 21$--$43$,
indicating
significant influence of nonlinear effects,
especially in the parts of the spectrum that
have wavevectors
highly oblique to the mean magnetic field,
as suggested in
Figs.~\ref{fig:sims} and \ref{fig:swdata}.

The expectation that
nonlinearities are strong in the solar wind as
kinetic scales are approached
is consistent with the detailed examples presented above, and
motivates a careful look, possibly on a case-by-case basis,
of the accuracy of wave damping and instability
computed from linear theory for application to
solar wind cascade, heating, and dissipation.
 It is noteworthy that the parameter
        $\Phi $
that we suggest here as an indicator of the expected influence of
cascade on linear kinetic processes is also
closely related to the controlling parameters identified
in studies of test particle energization
in various contexts,
including acceleration to high energies, anisotropic
generation of suprathermal particles, and plasma heating
        \citep{AmbrosianoEA88,%
                DmitrukEA03-testpart,DmitrukEA04-testpart,%
                ChandranEA10-perpHeat}.
Examination of
physics-based parameters of the type
given in Eq.~(\ref{TNLatdi})
may help to better understand
how the intensity of turbulence influences
the preferential absorption of energy into
proton thermal energy
as turbulence energy is increased, as recently
reported based on PIC simulation
        \citep{WuEA13-vKH}.

As a final remark, we note that the present discussion
has been based on a single nonlinear timescale that is
        \emph{characteristic}
of fluctuations at scale $1/k$.
Two types of complications enter immediately in any
more detailed treatments of timescales.
One is that the notion of locality in scale that we borrow
from Kolmogorov theory applies to shells in 
wavevector space---that is, 
we associate a single nonlinear timescale to
all wavevectors near a shell of wavenumber radius $k$.
Thus, formulas for estimating $\tauNL (k)$ such as
        Eq.~(\ref{tauNLk}) or (\ref{MHDtime})
involve the total energy near the shell,
that is the energy density integrated over a thin shell of
radius $k$.  This is a reasonable interpretation of
locality even when the distribution of energy over
the shell is anisotropic (see e.g., \cite{MattEA09-kspdiff}).
This necessarily involves averaging over regions (directions)
on the shell that may have very different energy levels and spectral
transfer properties.  Nonlocal transfer and its associated
timescales may be even more complex.
We avoid all such theoretical
complications here in an effort to elucidate the basic physical
timescale competition between linear and nonlinear effects.
Another underlying complication
is that the characteristic timescale is only an average
measure of the time variations at wavevector    $\vk $.
In reality we know that time variations are broadband
at each scale when nonlinear effects are strong.
Stated another way, in turbulence
the frequency ($ \omega$) spectra
        $ P( \vk, \omega) $ 
admit power over a broad range of frequency    
        $\omega$ 
for a given wavevector  $ \vk $,
as seen for example in a variety of
fluid, MHD, and plasma simulations
        \citep{DmitrukMatt07,
                DmitrukMatt09,ParasharEA10,
                VerscharenEA12,TenBargeHowes12,TenBargeEA13}.
The nonlinear timescale employed here is a standard estimate
of the average effect due to local couplings for fluctuations 
in the inertial range.
A more detailed treatment of the distribution of energy over
timescales (or frequencies)
would require examination of dynamical models
in more detail than is warranted here.
However, the present study may serve to
motivate future
more detailed studies of dynamical timescales
in plasma turbulence.

    \acknowledgments
This work
is supported in part by NSF (AGS-1063439,
AGS-1156094, SHINE), and by NASA
(NNX09AG31G, NNX11AJ44G, NNX13AD72G, MMS-IDS NNX08A083G,
MMS Theory and modeling Team, ISIS/Solar Probe Plus and the Heliospheric
Grand Challenges program ), 
Turboplasmas project (Marie Curie FP7 PIRSES-2010-269297),
POR Calabria FSE 2007/2013
and the UK STFC.
Some simulations were done on NCAR Yellowstone supercomputers, and on 
Blue Waters (NSF ACI 1238993).
 Hybrid Vlasov-Maxwell
simulations have been run within the European project PRACE Pra04-771.
%------------------------------------------------------------------

        \appendix
        \section{A: Alfv\'en crossing and nonlinear timescales}
        \label{app:A}

The Alfv\'en crossing time may be defined at the large scales
as
        $ \tau_A = L/V_A $,
for energy-containing scale $L$
and large-scale Alfv\'en speed $ V_A$
computed from the mean magnetic field
        $ \vB_0 $
as      $ V_A = B_0 /\sqrt{4 \pi \rho}$,
for a given mass density $\rho$.
By default we assume
the large scales to be isotropic but the definition is readily
generalized for imposed anisotropy.
Whenever
        $ V_A > Z \sim b$
for turbulence amplitude $Z$ and
        rms
magnetic fluctuation $ b$,
the ordering
        $ \tau_A < \tauNL $
holds.
However this timescale does
not influence
spectral transfer for strong turbulence,
for reasons discussed below.

The wavevector
dependent Alfv\'en time
        $ \tau_A( \vk ) = 1/ | \vk \cdot \vV_A |$
is just the reciprocal of the MHD
Alfv\'en wave frequency.
For normal modes with substantial
components
        $ k_\parallel$ parallel
to
        $ \vB_0$,
this timescale can be much shorter than the
corresponding nonlinear time
        $ \tauNL( \vk ) $
and can in principle influence spectral transfer
        \citep{PouquetEA76,ZhouEA04}.
However due to the usual dominant contribution of
        \emph{resonant triads}
to the nonlinear couplings
   \citep{ShebalinEA83,Grappin86,OughtonEA94},     % Bondeson85
the greatest contributions to nonlinear spectral
transfer are independent of
        $ \tau_A( \vk )$.
In this regard it is crucial to recall that the wavelike
couplings themselves make no contribution to
spectral transfer.    Rather the physics of
Alfv\'enic wavelike couplings may be understood as mainly
suppressing parallel spectral transfer, giving rise to spectral
anisotropy
        \citep{ShebalinEA83,CarboneVeltri90,OughtonEA94},
but generally  not having a major effect on the total rate at which
spectral transfer occurs.

For highly anisotropic turbulence, the role of the Alfv\'en time
may be of varying importance.
For the most anisotropic case---purely two-dimensional turbulence---the
large-scale Alfv\'en time
(computed in terms of the out of plane magnetic field)
does not contribute at all.
For low-frequency
        \emph{Reduced MHD},
the defining character of the dynamically important
region of wavevector space is simply that
        $ \tauNL( \vk) \leq \tau_A( \vk )$
        \citep{Mont82-strauss}.
For
        \emph{critical balance}
turbulence,
        $ \tauNL( \vk) \sim \tau_A( \vk) $,
which is usually interpreted as
        $ \tauNL( \vk) \approx \tau_A( \vk )$
        \citep{GoldreichSridhar95}.
The Alfv\'en timescale is not an independent
controlling factor for the rate of transfer in any of these cases.
Consequently in the analysis in the present paper, we focus exclusively
on the nonlinear timescale for comparisons with the linear Vlasov timescales
(where the Alfv\'en time again appears, but in connection with wave
behavior).
%---------------------------------------------------------------

        \section{B: Simulation Details}
        \label{app:B}

For the 2D MHD case, the 2D incompressible MHD equations are solved
in a $2\pi $-periodic box using a Fourier spectral method.
The simulation is a decaying run with initial kinetic and
magnetic energies equal to $0.5$, and initial energy excited within
a $ k$-band of
        $[5, 20]$.
The resolution of the simulation is
        $ 16384 \times 16384$,
with viscosity and resistivity
        $ \nu = \eta = 2.0\times 10^{-5}$
 \citep{WanEA13-xpts}.

The 3D incompressible Hall MHD simulation is also a free decay run
with initial kinetic and magnetic energies equal to 0.5
and an initially excited $k$-band of
        $ [2, 6] $.
A Fourier spectral method is employed in a 2$\pi$-periodic cube
with second-order Runge--Kutta timestepping.
The resolution is 512 modes in each direction,
with
        $ 1/ \di = 25 $,
and
        $ \nu = \eta = 3.0\times 10^{-3} $.

The 3D PIC simulation was performed using the general purpose PIC
plasma simulation code VPIC
        \citep{BowersEA08},
which solves
the relativistic Vlasov--Maxwell system of equations.
The initial
conditions correspond to uniform plasma with density $n_0$,
Maxwellian-distributed ions and electrons of equal temperature $T_0$,
a uniform magnetic field $B_0 \hat{\vct{z}}$,
and have
        $\beta_\text{p} = 0.5$.
The simulation domain is a cube of size $L \approx 41.9 \di $
with resolution of $2048^3$, such that the lowest
allowed wavelength in each direction is
        $ k_\text{min} \di = 0.15$.
The ion-to-electron mass ratio is $ m_i / m_e = 50$.
The turbulence is
seeded by imposing a perturbation of magnetic field initially,
with the two lowest modes in each direction initialized
        \citep{RoytershteynEA14}.

For the Vlasov simulation, the hybrid Vlasov--Maxwell equations
are solved using an Eulerian algorithm, in a five-dimensional geometry
(two dimensions in physical space and three in velocity space).
The 2D plane is perpendicular to the mean field $\vB_0$,
and fluctuating vectors have three components, in general.
The simulation is performed within a
        $ (2\pi\times 20 \di)^2 $ box,
with $512^2$ mesh points in space, and $51^3$
in the velocity space.
The initial condition consists of a Maxwellian plasma
perturbed by a 2D spectrum of Fourier modes, imposed for both the
velocity and the magnetic fields.
The plasma beta has been chosen equal to unity,
and the level of fluctuations is
        $ \delta b/ B_0 = 0.66 $.
More details can be found in
        \cite{ServidioEA12,ServidioEA14-vlasov}.

%---------------------------------------------------------------
% \section{bibliography}
%  \bibliographystyle{apj}
% \bibliography{ag,hl,mp,qz,refs_whm}  %%,TNL}

  \providecommand{\SortNoop}[1]{} %.......Use as {\SortNoop{Aaa}}
  \providecommand{\sortnoop}[1]{} %..............................
  \newcommand{\stereo} {\emph{{S}{T}{E}{R}{E}{O}}} %.................
  \newcommand{\au} {{A}{U}\ } %..................................
  \newcommand{\AU} {{A}{U}\ } %.................................
  \newcommand{\MHD} {{M}{H}{D}\ } %..............................
  \newcommand{\mhd} {{M}{H}{D}\ } %...............................
  \newcommand{\RMHD} {{R}{M}{H}{D}\ } %...........................
  \newcommand{\rmhd} {{R}{M}{H}{D}\ } %...........................
  \newcommand{\wkb} {{W}{K}{B}\ } %..............................
  \newcommand{\alfven} {{A}lfv{\'e}n\ } %...........................
  \newcommand{\alfvenic} {{A}lfv{\'e}nic\ } %.........................
  \newcommand{\Alfven} {{A}lfv{\'e}n\ } %...........................
  \newcommand{\Alfvenic} {{A}lfv{\'e}nic\ }

\end{document}